\documentclass[prl,twocolumn,showpacs,superscriptaddress,preprintnumbers]{revtex4-2}
\usepackage{amsmath,amssymb,amsfonts,float,graphics,epsfig,epstopdf,color,verbatim,tabularx,multirow,appendix}
\usepackage{bm}
\DeclareMathOperator{\Tr}{Tr}
\usepackage{graphicx}
\usepackage{xcolor}
\usepackage{dcolumn}
\usepackage{bm}
\usepackage{amsmath}
\usepackage{xspace}
\usepackage{time}

\usepackage{color}
\definecolor{LinkColor}{rgb}{0.256,0.439,0.588}
\usepackage{hyperref}
\hypersetup{
colorlinks=true,
citecolor=LinkColor,
linkcolor=LinkColor,
urlcolor=LinkColor
}
\newcommand{\ket}[1]{|#1\,\rangle}

\begin{document}
\title{Unconventional Scalings of Quantum Entropies in Long-Range Heisenberg Chains}

\author{Jiarui Zhao}
\affiliation{Department of Physics and HK Institute of Quantum Science \& Technology, The University of Hong Kong, Pokfulam Road, Hong Kong SAR}

\author{Nicolas Laflorencie}
\email{nicolas.laflorencie@cnrs.fr}
\affiliation{Laboratoire de Physique Th\'{e}orique, Universit\'{e} de Toulouse, CNRS, UPS, France}

\author{Zi Yang Meng}
\email{zymeng@hku.hk}
\affiliation{Department of Physics and HK Institute of Quantum Science \& Technology, The University of Hong Kong, Pokfulam Road, Hong Kong SAR}

\date{\today}

\begin{abstract}
In this work, building on state-of-the-art quantum Monte Carlo simulations, we perform systematic finite-size scaling of both entanglement and participation entropies for long-range Heisenberg chain with unfrustrated power-law decaying interactions. 
We find distinctive scaling behaviors for both quantum entropies in the various regimes explored by tuning the decay exponent $\alpha$, thus capturing non-trivial features through logarithmic terms, beyond the case of linear Nambu-Goldstone modes. 
Our systematic analysis reveals that the quantum entanglement information, hidden in the scaling of the two studied entropies, can be obtained to the same level of order parameters and other usual finite-size observables of quantum many-body lattice models. The analysis and results obtained here can readily apply to more quantum criticalities in 1D and 2D systems.
\end{abstract}

\maketitle
\noindent{\textcolor{blue}{\it Introduction.}---} Long-range interactions in quantum systems can give rise to unconventional quantum phases and transitions~\cite{duttaPhase2001,laflorencieCritical2005,2007arXiv0709.4487B,sandvikGround2010,knapProbing2013,feyCritical2016,leporiEffective2016,maghrebiContinuous2017,liLong2021,songQuantum2023,zhaoFinite2023,songDynamical2023,diesselGeneralized2023,liaoCaution2023,wangValidity2023,defenuLongrange2023,leeLandau2023}. Particularly interesting is the case of low spatial dimensions $d=1,2$, where unusual behaviors can occur beyond the realm of Hohenberg-Mermin-Wagner theorem~\cite{BrunoAbsence2001,merinAbsence1966,hohenbergExistence1967,FisherCritical1972,SakRecursion1973}, such as long-range order that breaks the continuous symmetry in the ground state for $d=1$~\cite{laflorencieCritical2005}, the modification of the excitation spectra~\cite{yusufSpin2004,diesselGeneralized2023,songDynamical2023,defenuLongrange2023,adelhardtMonte2024}, the breaking of the Lieb-Robinson bound of propagation of information~\cite{frerotMultispeed2018,vanderstraetenQuasiparticle2018,tranLocality2019,colmenarezLieb2020} and even the violation of area-law scaling
of entanglement entropy (EE)~\cite{koffelEntanglement2012,frerotEntanglement2017,liLong2021}. 
\begin{figure}[t]
\includegraphics[width=\columnwidth]{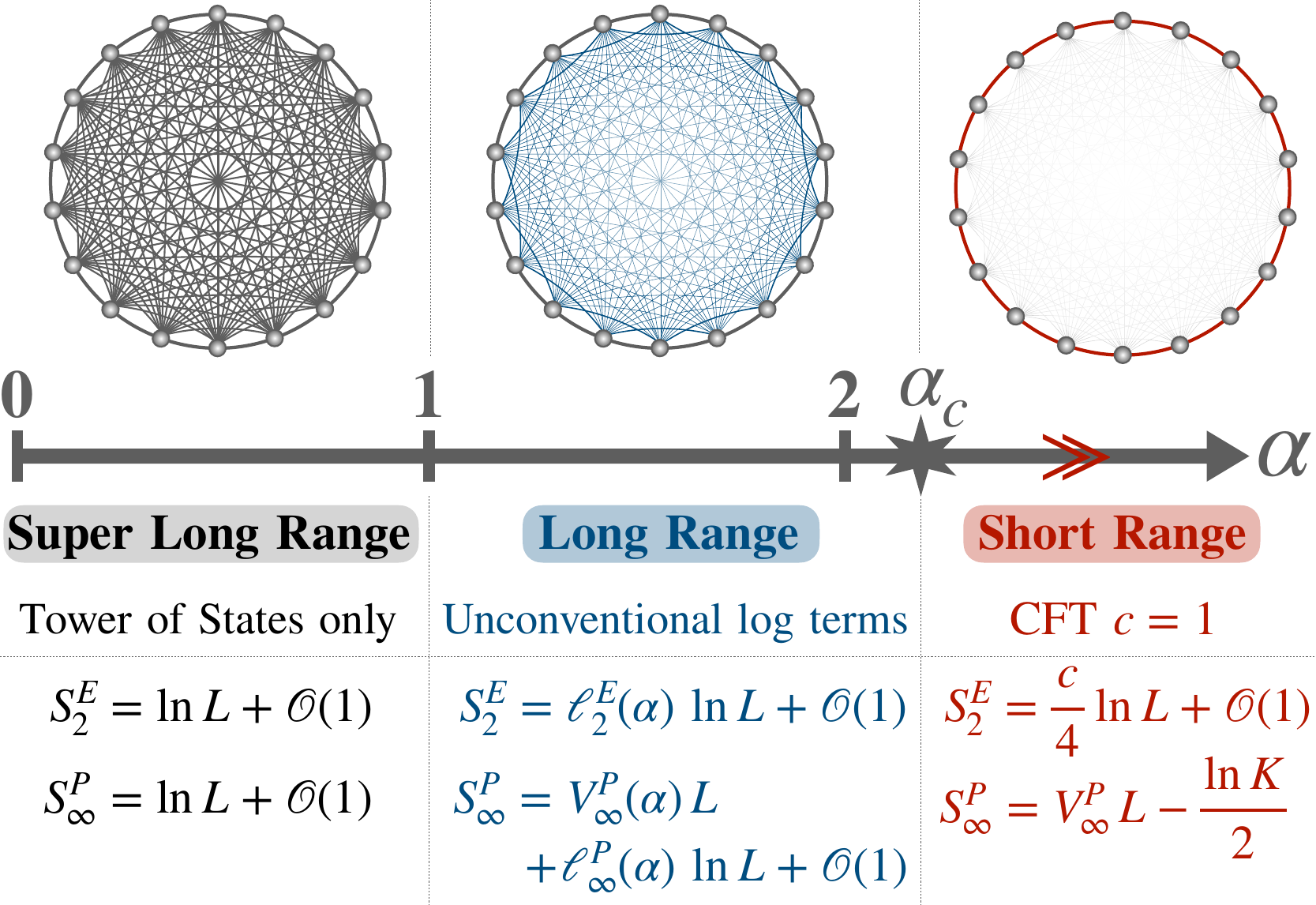}
\caption{Schematic picture showing the succession of the three different regimes as function of the exponent $\alpha$ of the long-range couplings in Eq.~\eqref{eq:H_LR}. Left: for $\alpha\in [0,1]$, the system is over-extensive and falls in the fully connected Lieb-Mattis case~\cite{lieb1962ordering}. Right: for $\alpha> 
\alpha_c\approx 2.23$, long-range couplings are irrelevant~\cite{laflorencieCritical2005} (signaled by the red arrows) and the quasi-long-range-ordered Luttinger liquid physics is recovered with the CFT results for both EE and PE. Middle: for $1<\alpha<\alpha_c$, the ground-state has a true SU(2) symmetry broken long-range order, and displays finite unconventional logarithmic terms for both EE and PE. $\alpha_c$ is the unconventional QCP also with logarithmic corrections in the two entropies. See text for more details on the various scaling forms.}
\label{fig:sketch}
\end{figure}
However, these exotic features in long-range systems have been discussed on a case-by-case basis, and there is a lack of systematic study of the quantum critical behavior and the scaling of the entanglement content as the interaction continuously changes from long-range to short-range.
In such a case, one would like to draw a comprehensive picture which combines critical behavior with the scalings of both EE~\cite{calabreseEntanglement2004,laflorencieQuantum2016} and the participation entropy (PE)~\cite{stephanShannon2009,stephanRenyi2010,zaletelLogarithmic2011}.  

Introduced in Refs.~\cite{yusufSpin2004,laflorencieCritical2005}, the unfrustrated long-range antiferromagnetic (AF) Heisenberg $S=1/2$ chain model
\begin{equation}
{\cal{H}}_{\rm LR}(\alpha)=\sum_{i<j}J_{ij} \mathbf{S}_{i}\cdot \mathbf{S}_{j},\quad J_{ij}=\frac{(-1)^{j-i+1}}{|i-j|^{\alpha}},
\label{eq:H_LR}
\end{equation}
 provides an ideal test-bed to explore a very rich variety of non-trivial 
phenomena~\cite{laflorencieCritical2005,2007arXiv0709.4487B,sandvikGround2010,yangFrom2021}. Upon tuning the decay exponent $\alpha$, there is an unconventional quantum critical point (QCP) at $\alpha_c\sim 2.2-2.25$~\cite{laflorencieCritical2005,2007arXiv0709.4487B,sandvikGround2010}  between a short-range (SR) regime described by conformal field theory (CFT) for $\alpha>\alpha_c$, and a long-range (LR) ordered phase for $1<\alpha<\alpha_c$ where the continuous SU(2) symmetry spontaneously breaks, even if we are in $d=1$. 
It is also worth mentioning the "over-extensive" case $\alpha\le 1$~\cite{botzungEffects2021} (super-long-range), where  quantum corrections to the classical AF order parameter vanish at large sizes~\cite{yusufSpin2004}, putting this regime in the same class as the fully connected Lieb-Mattis model~\cite{lieb1962ordering}, as will be discussed below.

In this work, building on both entanglement and participation entropies obtained from large-scale quantum Monte Carlo simulations~\cite{Sandvik1999,sandvikQuantum2002,sandvikStochastic2003}, we explore how the critical properties, the entanglement content and the complexity of the many-body ground-state evolves upon changing the exponent $\alpha$ of the LR Heisenberg interaction in Eq.~\eqref{eq:H_LR}. As summarized in Fig.~\ref{fig:sketch} three situations emerge, with characteristically different scalings for the two quantum entropies. While both short-range (SR) and super-long-range (super-LR) regimes can be well understood from current theoretical frameworks, the broad intermediate long-range (LR) ordered phase is much more intriguing. Indeed, in this long-range regime, and at the unconventional critical point $\alpha_c$, where Lorentz invariance is broken (i.e. the dynamical critical exponent $z<1$~\cite{laflorencieCritical2005,sandvikGround2010}), non-trivial logarithmic scalings are found for both quantum entropies EE and PE.

The rest of this Letter is organized as follows. We first revisit the critical point with standard finite-size scaling of the Binder cumulant, which allows us to locate the critical point with higher accuracy.
We then discuss our QMC results for the two quantum entropies (first the EE and then the PE) over the whole regime. We clearly confirm the scaling forms for both SR and super-LR phases, and discuss in detail the extraction of the unconventional logarithmic scalings in the intermediate LR regime. Finally, we discuss the implications of our results and possible experimental consequences.

\noindent{\textcolor{blue}{\it Model and Unconventional Critical Point.}---} 
The 1D LR Heisenberg model we study is defined in Eq.~\eqref{eq:H_LR},
where $J_{ij}$ represents the staggered long-range interaction which does not introduce frustrations. In practice, in order to alleviate finite-size effects, we consider the Ewald summation which modifies the couplings strength as $\tilde{J}_{ij}=\sum_{m=-\infty}^{+\infty} \frac{(-1)^{i-j+1}}{|i-j+mL|^{\alpha}}$, as has been successfully applied in long-range spin models~\cite{Flores-Sola2015,FukuiOrder2009,KoziolQuantum2021,AdelhardtContinuously2022,songQuantum2023,zhaoFinite2023,songExtracting2023,liaoExtracting2024}. As discussed in Ref.~\cite{laflorencieCritical2005}, the model exhibits a true long-range order if $\alpha< \alpha_c$ and quasi-long-range order when $\alpha>\alpha_c$. In between, at $\alpha_c=2.225(25)$ there is a peculiar quantum critical point with a dynamical exponent $z<1$~\cite{laflorencieCritical2005,sandvikGround2010}. The Hamiltonian Eq.~\eqref{eq:H_LR} is sign-problem-free, and we use the stochastic series expansion (SSE) quantum Monte Carlo (QMC) method~\cite{Sandvik1999,sandvikQuantum2002,sandvikStochastic2003} to simulate this model. We measure the R\'enyi entanglement entropy (EE) with the help of nonequilibrium increment method~\cite{albaOutofequilibrium2017,demidioEntanglement2020,zhaoMeasuring2022,zhaoScaling2022}, and also compute the participation entropy (PE)~\cite{stephanShannon2009,stephanRenyi2010,zaletelLogarithmic2011,stephanPhase2011,atasMultifractality2012,alcarazUniversal2013,stephanShannon2014,luitzUniversal2014} during the SSE simulations.

\begin{figure}[htp!]
\includegraphics[width=\columnwidth]{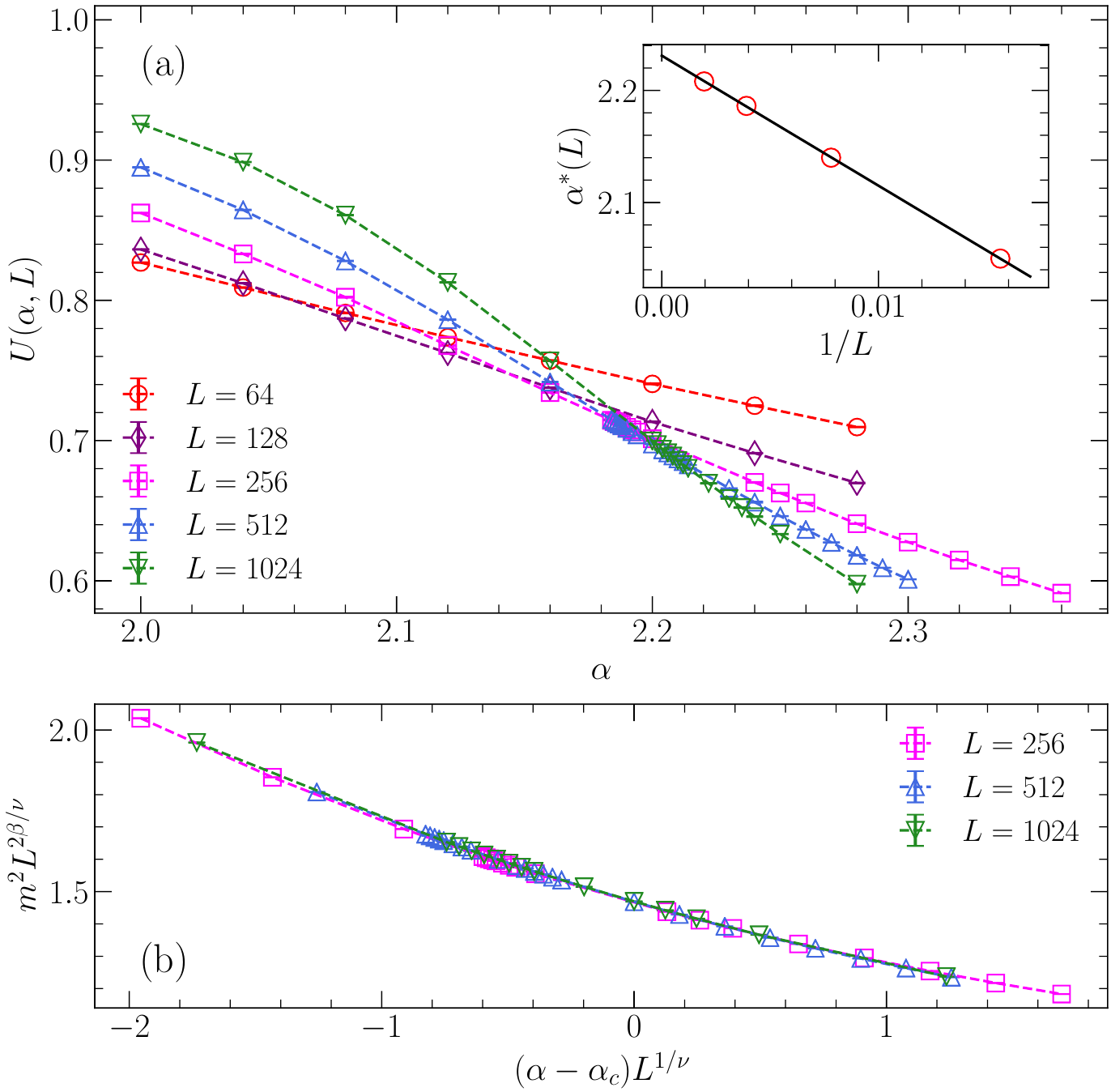}
\caption{Determination of the critical point $\alpha_c$ with Binder cumulants and data collapse of order parameter. (a) Binder cumulants $U(\alpha,L)$ versus $\alpha$ for different system sizes. (Inset) The crossing point of $U(\alpha,L)$ with $U(\alpha,2L)$ versus $1/L$, from which the extrapolation $\alpha_c=2.2307(9)$ is determined. (b) Data collapse of the N\'eel order parameter close to $\alpha_c$, from which the optimised $\beta=0.57(2)$ and $\nu=2.16(1)$ are determined.}
\label{fig:fig2}
\end{figure}

We first determine the critical point accurately by considering the Binder cumulants defined as
 $   U = 1 - \frac{\langle m^4 \rangle}{3 \langle m^2 \rangle^2}$,
where $m=\frac{1}{N}\sum_{i=1}^{N}(-1)^{i}S_{i}^{z}$ is the N\'eel order parameter. The Binder cumulant is a dimensionless quantity, and according to the finite-size crossing point analysis~\cite{qinDuality2017}, the crossing points ($\alpha^*(L)$) of $U(\alpha,L)$ and $U(\alpha,2L)$ is expected to converge to the phase transition point ($\alpha_c$) by $\alpha^*(L)=\alpha_c+L^{-1/\nu+\omega}$ where $\nu$ is the universal correlation length exponent and $\omega$ is a non-universal correction exponent which is related with the leading irrelevant field in the renormalization flow. We thus measure the Binder cumulants for system sizes $L=64,128,256,512,1024$ and fix the inverse temperature $\beta=\frac{1}{T}=L$ in the simulation. The crossing point is determined by fitting the data set with a cubic polynomial function and solving the intersection point of the fitted curves. This procedure is visible in Fig.~\ref{fig:fig2} where panel (a) shows the Binder cumulant of the AF order for different system sizes as a function of $\alpha$. In the inset of Fig.~\ref{fig:fig2} (a), from the extrapolation of power-law fitting of the crossing points to the thermodynamic limit, we obtain the critical point at $\alpha_c=2.2307(9)$, consistent and yet with higher accuracy compared with Refs.~\cite{laflorencieCritical2005,2007arXiv0709.4487B,sandvikGround2010}. With such $\alpha_c$, we further collapse the AF order parameter $m^2$ to determine the critical exponents with the finite size scaling relation
 $   m^2 \cdot L^{2\beta/\nu}= f\left[(\alpha-\alpha_c)\cdot L^{1/\nu}\right]$,
where $\beta$ and $\nu$ are separately the critical exponents associated with order parameter and correlation length. As shown in Fig.~\ref{fig:fig2} (b), we fix $\alpha_c=2.23$ and adjust the values of $\beta$ and $\nu$ to obtain the set of $(\beta,\nu)$ which collapse the data the best. Through this method, we obtain $\beta=0.57(2)$ and $\nu=2.16(1)$ are critical exponents at this transition point, which are consistent with Refs.~\cite{laflorencieCritical2005,2007arXiv0709.4487B}. Together with the fact that the dynamic exponent $z<1$~\cite{laflorencieCritical2005,sandvikGround2010}, this is a new quantum critical point distinct from conventional (1+1) dimensional QCPs, and we will study its nature through scaling of both EE and PE below.\\

\begin{figure}[htp!]
\includegraphics[width=\columnwidth]{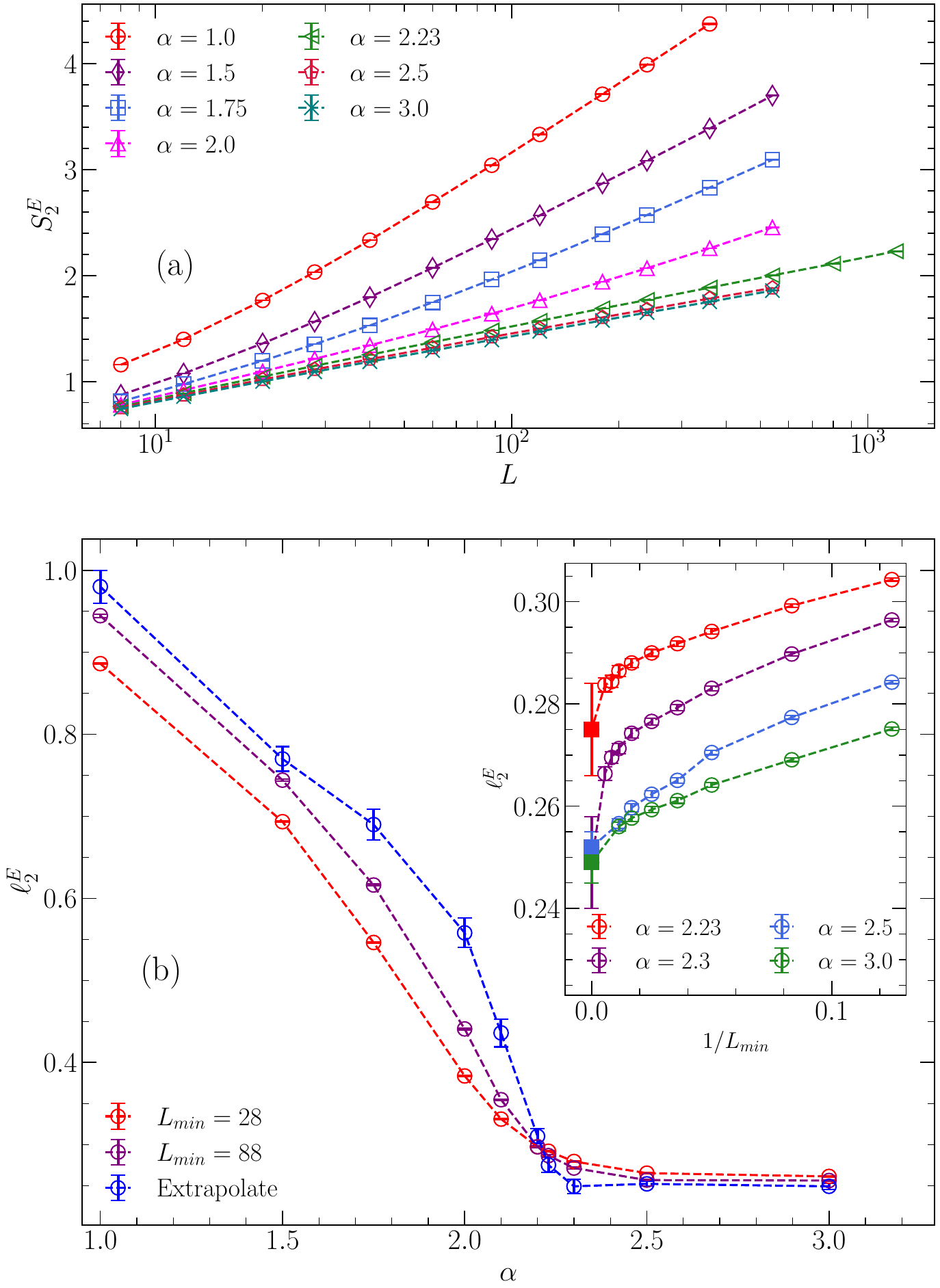}
\caption{Unconventional scaling of the half-chain R\'enyi EE for ${\cal{H}}_{\rm LR}(\alpha)$ Eq.~\eqref{eq:H_LR}. (a) $S^{E}_{2}$ versus $L$ for different $\alpha$. (b) Extrapolation of fitted log-coefficient $\ell^{E}_{2}(\alpha)$ to the thermodynamic limit. In the analysis we gradually slide the fitting window which contains 6 data points {for $\alpha>1$ and 5 data points for $\alpha=1$}, and $L_{\text{min}}$ is the smallest system size in the fitting window. Inset exemplifies the finite-size and extrapolated $\ell^{E}_{2}(\alpha)$ for different $\alpha$ close to $\alpha_c$. For $\alpha> \alpha_c$, the system is described by a short-range (1+1)D CFT with $\ell^{E}_{2}(\alpha)=\frac{1}{4}$; for $\alpha \leq 1$, the system is in super LR regime with $\ell^{E}_{2}(\alpha)=1$. At $\alpha_c$, we find the $\ell^{E}_{2}(\alpha)=0.275(9)$ and it continuously increases in the long-range regime to $\ell^{E}_{2}(\alpha)=1$ in the super LR case.}
\label{fig:fig3}
\end{figure}

\noindent{\textcolor{blue}{\it Entanglement Entropy}---} 
The R\'enyi EE defined as $S_{n}^{E}=\frac{1}{1-n}\ln \Tr \rho_A^{n}$ ($\rho_A$ is the reduced density matrix of subsystem $A$), can be equivalently written as $S_{n}^{E}=\frac{1}{1-n}\ln \frac{Z_{A}^{(n)}}{Z^{(n)}}$ based on its trace structure~\cite{calabreseEntanglement2004,laflorencieQuantum2016}.  In SSE QMC's configuration space~\cite{Sandvik1999}, $Z_{A}^{(n)}$ can be represented as $n$ replicas of space-time configurations with region A glued together in imaginary time, and $Z^{(n)}$ represents n independent replicas. For a (1+1)d conformal field theory (CFT), if we fix region $A$ to be half of the chain, R\'enyi EE is analytically predicted to scale as~\cite{calabreseEntanglement2004}
\begin{equation}
\label{eq:cft}
    S_{n}^{E}=\frac{c}{6}\left(1+\frac{1}{n}\right)\ln L+O(1),
\end{equation}
where $c$ is the central charge of the CFT. For $\alpha>\alpha_c$, the LR interaction is irrelevant~\cite{laflorencieCritical2005} so that we expect the EE to scale as Eq.~\eqref{eq:cft}. In the $n=2$ case, $S^{E}_2=\frac{c}{4}\ln L + O(1)$ with $c=1$ for short-range $SU(2)$ Heisenberg chain~\cite{affleckUniversal1986}. At the QCP $\alpha=\alpha_c$ and inside the LR regime $1<\alpha<\alpha_c$, the system is not described by a CFT, and the exact scaling of EE is not fully understood, despite very inspiring semi-classical results~\cite{frerotEntanglement2017}. 
As detailed below, we find that our QMC data is perfectly described by the following logarithmic scaling 
\begin{equation}
    S^{E}_2 = \ell^{E}_{2}(\alpha)\ln L + O(1),
    \label{eq:log}
\end{equation} 
with
an unconventional $\alpha$-dependent coefficient $\ell^{E}_{2}(\alpha)$. In the LR regime where the system breaks SU$(2)$ symmetry,
$\ell^{E}_{2}(\alpha)$ is finite, and continuously increases from $\ell^{E}_{2}(\alpha_c)=0.275(9)$  to $\ell^{E}_{2}(\alpha=1)\to 1$. In the super-LR regime, the EE follows $S_2^{E} = \ln L + O(1)$, due to the tower of states (TOS) structure~\cite{vidalEntanglement2007}.

We compute the 2nd R\'enyi EE $S^{E}_2$ across the phase transition using the nonequilibrium increment method developed very recently and has proven to be efficient in $2d$ and $(2+1)d$ quantum spin systems~\cite{albaOutofequilibrium2017,demidioEntanglement2020,zhaoMeasuring2022,zhaoScaling2022,songExtracting2023,liaoExtracting2024}. The basic idea behind this method is to regard  $S_{n}^{E}$ with the free energy difference between $Z_{A}^{(n)}$ and ${Z^{(n)}}$ by $S_{n}^{E}=\frac{1}{1-n}\ln \frac{Z_{A}^{(n)}}{Z^{(n)}}=\frac{\beta}{n-1}\Delta F$, and relate it with the ensemble average of the total work done during many nonequlibrium evolution processes between these two systems. According to Jarzynski's equality~\cite{Jarzynski1997} from statistic physics, if one system is gradually evolved from another via tunning its external parameters, although the total work done ($W$) during such process is always not less than the free energy difference($\Delta F$), they are strictly related with $\Delta F=-\beta^{-1}\ln \overline{e^{-\beta W}}$, where the average is taken over different tunning processes. The R\'enyi entanglement entropy is then calculated by designing such tunning processes from $Z_{A}^{(n)}$ to ${Z^{(n)}}$ and accumulating the total work done during this process. The detailed implementation of this method and its optimizations are documented in Refs.~\cite{albaOutofequilibrium2017,demidioEntanglement2020,zhaoMeasuring2022,zhaoScaling2022}.

Fig.~\ref{fig:fig3} (a) shows our results of EE for different $\alpha\in [1,\,3]$ at system size $L_{\text{max}}=540$ for $\alpha>1$, and $L_{\text{max}}=360$ for $\alpha=1$. As the system becomes more and more long-ranged, the finite-size effects manifest. We fit our results with Eq.~\eqref{eq:log} and fix the number of data points in fitting window to be 6. In this case, we gradually slide the fitting window and the smallest system size in the window is denoted by $L_{\text{min}}$. As exemplified in the inset of Fig.~\ref{fig:fig3} (b), we use a quadratic function to fit the data points and extrapolate the log-coefficient $\ell^{E}_{2}(\alpha)$ to the thermodynamic limit ($1/L_{\text{min}}\rightarrow 0$). At the critical point, the data points extrapolate to $\ell^{E}_{2}(\alpha_c)=0.275(9)$ which is intrinsically different from the SR CFT case $\alpha > \alpha_c$ with $\ell^{E}_{2}(\alpha)=0.25$. The main panel of Fig.~\ref{fig:fig3} (b) thus shows the extrapolated $\ell^{E}_{2}(\alpha)$ with increasing $L_{\rm min}$ in the whole regime of $\alpha$. We observe that when $\alpha=1$, the extrapolated $\ell^{E}_{2}(\alpha)=0.98(2)$, consistent with the expectation of the super-LR regime. As $\alpha$ grows, $\ell^{E}_{2}(\alpha)$ gradually decreases and only gets to its CFT value $1/4$ when $\alpha>\alpha_c$.  Note that close to $\alpha_c=2.23$ but in the SR regime, the extrapolation procedure yields a slightly larger value of $\ell^{E}_{2}(\alpha)$ than the expected $0.25$. We attribute this to strong finite-size effects to large correlation length in the critical regime. Besides the extrapolation shown in the inset of Fig.~\ref{fig:fig3}(b) for a few $\alpha$ values, the SM~\cite{suppl} shows extrapolation process of $\ell^{E}_{2}(\alpha)$ for the other $\alpha$.\\

\noindent{\textcolor{blue}{\it Participation Entropy.}---} Another way to probe the quantum complexity of a many-body wavefunction is based on the participation entropy (PE), a quantity that has been shown to be very useful in capturing the universality of different quantum phases~\cite{stephanShannon2009,stephanRenyi2010,zaletelLogarithmic2011,stephanPhase2011,atasMultifractality2012,alcarazUniversal2013,stephanShannon2014,luitzUniversal2014,luitzParticipation2014,maceMultifractal2019,sierantUniversal2022}. In short, for a given quantum state ${\ket{\Psi}}=\sum_j a_j{\ket{j}}$, expanded in a computational basis $\{\ket{j}\}$, one can naturally interpret $p_j=|a_j|^2$ as the probability to occupy the configuration $j$ by the  normalized state ${\ket{\Psi}}$. One can then build the R\'enyi PEs (for a review, see \cite{luitzParticipation2014}) defined as follows
\begin{equation}
S_{n}^{P}=\frac{1}{1-n}\ln\sum_j\left(p_j\right)^n.
\end{equation}
Below, we only focus on the $n=\infty$ case, which is the easiest limit to handle with QMC since it only requires to record the most frequent basis state during the SSE sampling~\cite{luitzUniversal2014}, yielding $S_{\infty}^{P}=-\ln p_{\rm max}$. For the AF model Eq.~\eqref{eq:H_LR}, the most probable spin configurations in the $\{S^z\}$ basis are the two (degenerate) N\'eel states: $\ket{\hskip -0.1cm\uparrow\downarrow\uparrow\downarrow\ldots}$ and $\ket{\hskip -0.1cm\downarrow\uparrow \downarrow\uparrow\ldots}$.
It is remarkable that the knowledge of a single coefficient $p_{\rm max}$ (out of an exponentially large number) is sufficient to capture non-trivial quantum criticality ~\cite{stephanShannon2009,stephanRenyi2010,zaletelLogarithmic2011} or broken symmetry states~\cite{luitzUniversal2014,misguichFinite2017,luitzQuantum2017}.
\begin{figure}[htp!]
\includegraphics[width=\columnwidth]{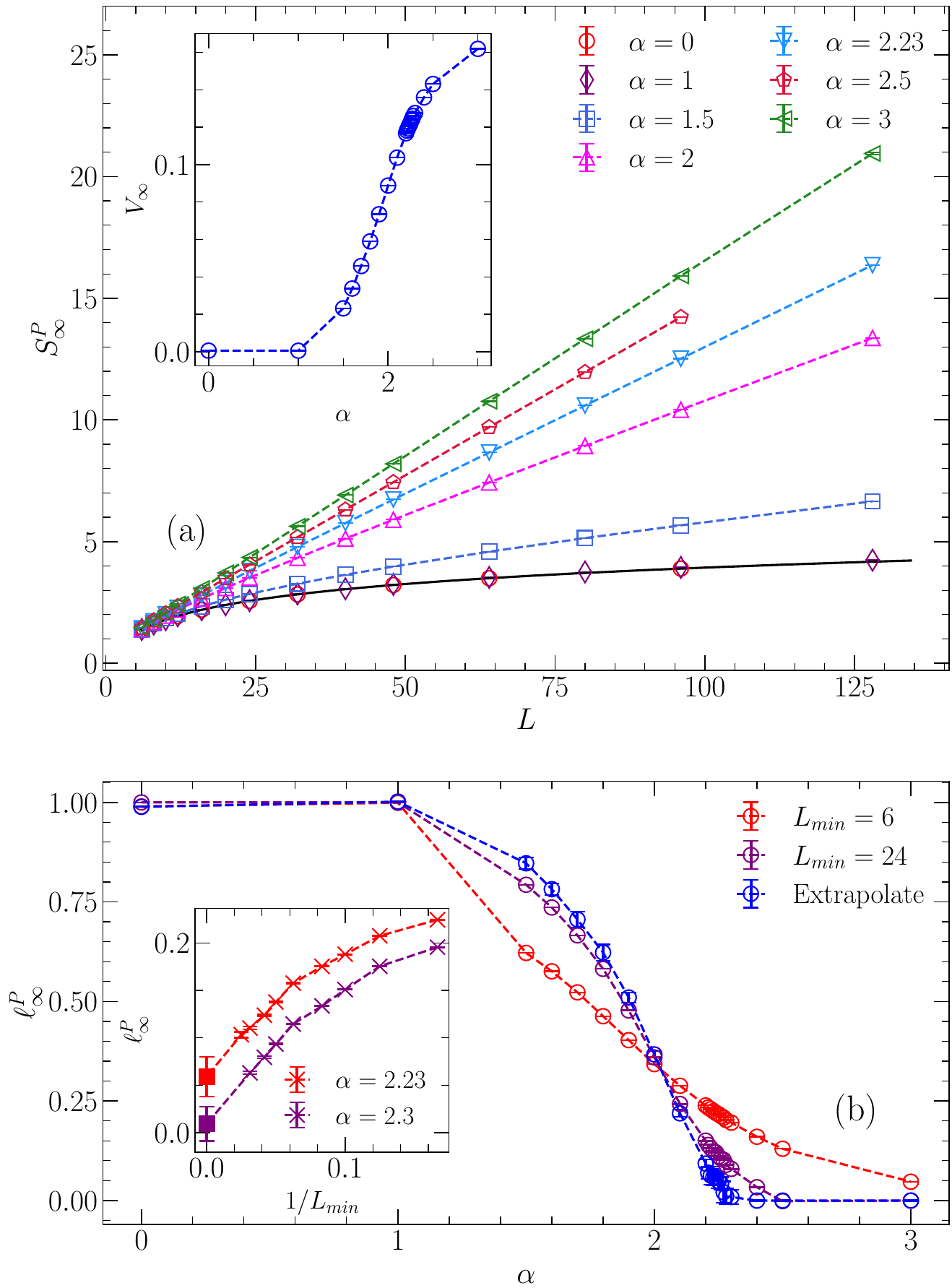}
\caption{Unconventional scaling of the PE for ${\cal{H}}_{\rm LR}(\alpha)$ Eq.~\eqref{eq:H_LR}. (a) $S^{{P}}_{\infty}$ versus system size $L$ for different $\alpha$. Inset shows the extrapolated volume-law coefficient $V_{\alpha}$ for different $\alpha$. {The solid black curve represents the relation of $S^{P}_{\infty}=\ln(L/2+1)$, which matches perfectly with finite-size PE data for both $\alpha=0$ and $\alpha=1$.} (b) Extrapolation of fitted log-coefficient $\ell^{P}_{\infty}(\alpha)$ to the thermodynamic limit. In the analysis we gradually slide the fitting window which contains 5 data points for $\alpha\le1$ and 6 data points for other $\alpha$, and $L_{\text{min}}$ is the smallest system size in the fitting window. Inset shows the finite size extrapolation of log-coefficient $\ell^{P}_{\infty}(\alpha)$ for $\alpha_c$ and $\alpha=2.3$. $\ell^{P}_{\infty}(\alpha_c)=0.059(21)$   yields an unconventional value for critical point and $\ell^{P}_{\infty}(\alpha=2.3)=0.009(18)$ in the short-range regime. }
\label{fig:fig4}
\end{figure}
For continuous symmetry breaking, the finite-size scaling of the PE has been first found to obey~\cite{luitzUniversal2014}
\begin{equation}
    S_{\infty}^{P}(L)=V_{\infty}(\alpha)L+\ell^{P}_{\infty}(\alpha)\ln L+O(1),
    \label{eq:eq6}
\end{equation}
with universal subleading logarithmic corrections such that $\ell^{P}_{\infty}(\alpha)$ directly depends on the number of linearly dispersing Nambu-Goldstone bosons $n_{\rm NG}$. The original conjecture $\ell^{P}_{\infty}(\alpha)=n_{\rm NG}/4$, made from high-precision QMC data on the 2D square lattice~\cite{luitzUniversal2014},  was later shown analytically  by Misguich, Pasquier and Oshikawa~\cite{misguichFinite2017} who found two contributions of opposite signs, one (positive) coming form the TOS and the other (negative) from the spin-waves (SW). Below we aim at extending this to the peculiar case of broken continuous symmetry in $d=1$, with sublinear dispersing SW~\cite{yusufSpin2004,laflorencieCritical2005,frerotEntanglement2017}.

For this purpose, ${\cal{H}}_{\rm LR}(\alpha)$ Eq.~\eqref{eq:H_LR} is very instructive, as it interpolates between two well-known limits. (i) In the SR regime for $\alpha>\alpha_c$, 
the Luttinger liquid (LL) behavior is expected~\cite{stephanShannon2009,stephanPhase2011} with  $S_{\infty}^{P}=V_{\infty}(\alpha)L-\ln\sqrt{K}+o(1)$, where $K$ is the LL parameter and $V_{\alpha}$ is a non-universal volume-law term that encodes the generic ground-state multifractality~\cite{atasMultifractality2012,luitzUniversal2014}. (ii) The opposite limit $\alpha\to 0$ describes the fully connected Lieb-Mattis (LM) model~\cite{lieb1962ordering}, where the exact ground-state wave function is known~\cite{vidalEntanglement2007}, yielding $S_{\infty}^{P}=\ln\left(L/2+1\right)$~\cite{luitzUniversal2014}. It is also natural to expect LM physics as soon as $\alpha< 1$~\cite{noteLM}. On the other hand, in the long-range regime $1<\alpha<\alpha_c$ and at the QCP $\alpha_c$, we find that PE follows the unconventional scaling form of Eq.~\eqref{eq:eq6} with a continuously varying $\ell^{P}_{\infty}(\alpha)$.

Our results of $S_{\infty}^{P}(\alpha,L)$ are shown in Fig.~\ref{fig:fig4}. First, panel (a) clearly shows that in the super LR case for $\alpha=0,1$, QMC data follow the $S_{\infty}^{P}=\ln\left(L/2+1\right)$ behavior, with $V_{\infty}(\alpha)=0$ as denoted in the inset. Inside the LR regime and at the QCP, the generic scaling Eq.~\eqref{eq:eq6} is observed, with unconventional logarithmic corrections and varying $\ell^{P}_{\infty}(\alpha)$. Further increase $\alpha$ to $\alpha >\alpha_c$, the volume-law becomes dominant and the log-coefficient vanishes $\ell^{P}_{\infty}(\alpha)=0$, thus recovering the LL behavior. Fig.~\ref{fig:fig4} (b) exhibits such evolution, the inset focusing on the log-coefficient close to $\alpha_c$. As $\alpha$ varies from 0 to large values, the log-coefficient $\ell^{P}_{\infty}(\alpha)$ in $S_{\infty}^{P}$ changes from 1 inside the super LR regime $\alpha\le1$, to a finite non-trivial value that should be related to the TOS and sublinear dispersing SW in the LR regime $1<\alpha<\alpha_c$. We remarkably observed a small but finite critical value $\ell^{P}_{\alpha=\alpha_c}\approx 0.059(18)$, which presumably jumps to 0 in the SR regime when $\alpha > \alpha_c$. The detailed fitting procedures are documented in the SM~\cite{suppl}. The existence of unconventional logarithmic correction to the PE, inside the long-range regime and at the QCP, further reinforces the consistent picture obtained from the scaling of EE in Fig.~\ref{fig:fig3}.

\begin{figure}[b!]
\includegraphics[width=\columnwidth]{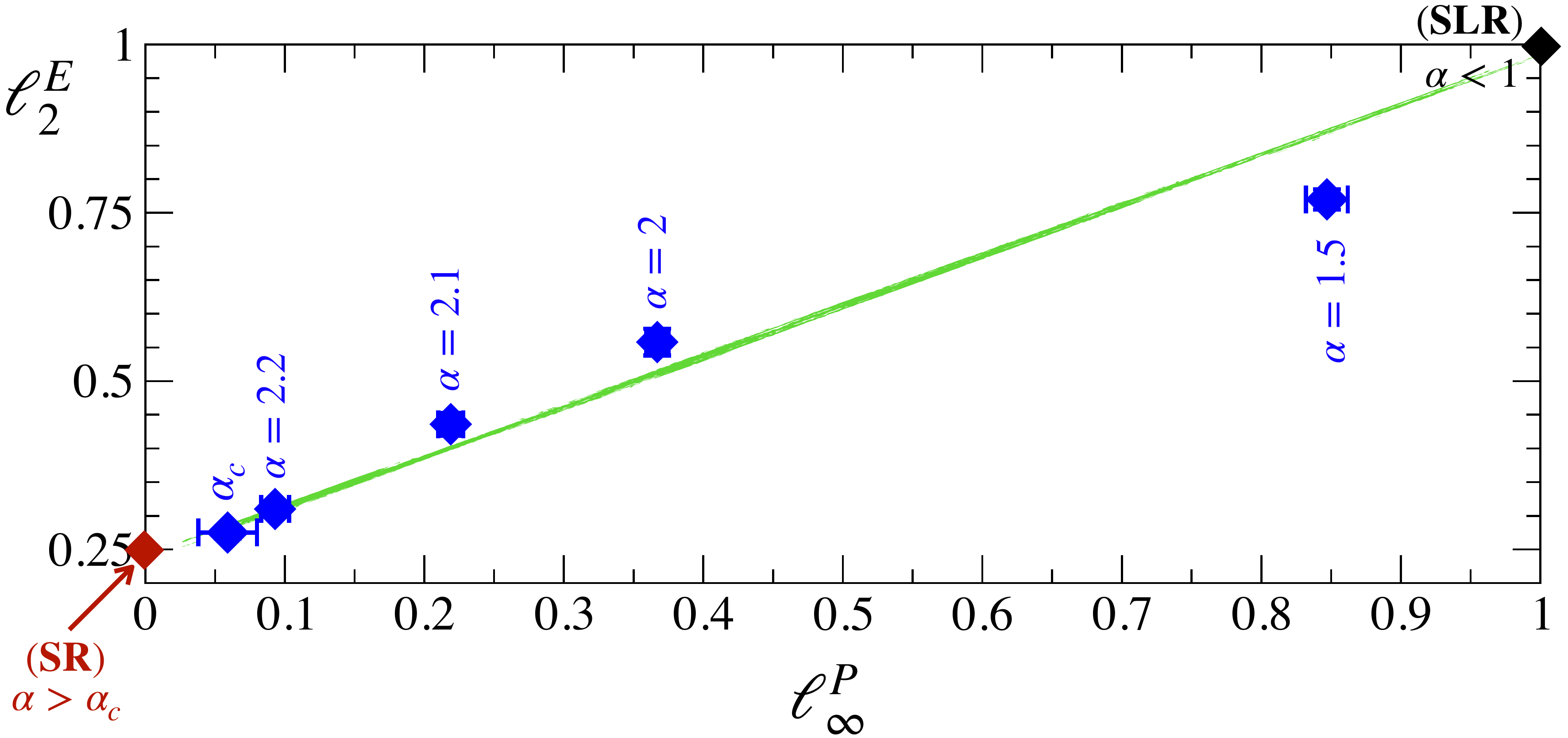}
\caption{Relationship between $\ell^P_{\infty}(\alpha)$ and $\ell^E_2(\alpha)$ in the critical and LR regimes (the green line is a guide to the eyes).  In the SR regime, ($\ell^{P}_{\infty}=0,\ell^E_2=\frac{1}{4})$ as denoted by the red diamond. In the super-LR regime, ($\ell^{P}_{\infty}=1,\ell^E_2=1)$ as denoted by the black diamond.}
\label{fig:fig5}
\end{figure}

\noindent{\textcolor{blue}{\it Discussions.}}--- We also find an interesting relation between the log-coefficients of the two entropies $\ell^E_2$ and $\ell^P_{\infty}$, as shown in Fig.~\ref{fig:fig5}. Starting from the SR regime where $\ell^{P}_{\infty}=0$ and $\ell^E_2=\frac{1}{4}$, the obtained estimates for $\ell^P_{\infty}(\alpha)$ and $\ell^E_2(\alpha)$ appear to be somehow related. It is tempting to interpret this observed linear dependence in the QCP and LR regimes as the result of a non-trivial combination of TOS and SW contributions~\cite{metlitskiEntanglement2011,songEntanglement2011,frerotEntanglement2017,misguichFinite2017}, but we leave this intriguing effect for further investigation.

To summarize, we have performed systematic finite size scaling of the entanglement entropy (EE) and participation entropy (PE) for long-range Heisenberg chain. We find distinctive scaling behaviors of both quantum entropies in the Super-LR $\alpha \le 1$, LR $1<\alpha<\alpha_c\approx2.23$, at the QCP $\alpha_c$ and inside the SR $\alpha>\alpha_c$ regimes. It is interesting that both entropies successfully reveal the unconventional logarithmic terms signifying the existence of the long-range order inside the LR regime, and at the QCP, separating the LR and SR regimes. The $\ell^{E}_{\alpha}$ and $\ell^{P}_{\infty}(\alpha)$ obtained also behave correctly inside the Super-LR and SR regimes where the analytic forms of EE and PE are known due to the fully connected Lieb-Mattis model~\cite{lieb1962ordering} and exact knowledge of the ground state wavefunction of SU(2) Heisenberg chain with short-range interaction. 

Our results and systematic analysis reveal that the quantum entanglement information, hidden in the finite size scaling of the two entropies employed here, can be obtained to the same level of order parameters and other usual finite size observables in the quantum many-body lattice models. We also foresee the experimental verification of our results, as quite remarkably, continuous symmetry breaking in spin chains with long-range interactions has recently been realized in trapped-ion quantum simulator~\cite{fengContinuous2023}. 



\begin{acknowledgments}
{\it Acknowledgments}\,---\,  JRZ and ZYM acknowledge the support from the Research Grants Council (RGC) of Hong Kong Special Administrative Region of China (Project Nos. AoE/P-701/20, 17309822, HKU C7037-22GF, 17302223), the ANR/RGC Joint Research Scheme sponsored by RGC of Hong Kong and French National Research Agency (Project No. A\_HKU703/22). We thank HPC2021 system under the Information Technology Services and the Blackbody HPC system at the Department of Physics, University of Hong Kong, as well as the Beijng PARATERA Tech CO.,Ltd. (URL: https://cloud.paratera.com) for providing HPC resources that have contributed to the research results reported within this paper. NL acknowledges the use of HPC resources from CALMIP (grants 2022-P0677 and 2023-P0677) and GENCI (projects A0130500225 and A0150500225).
\end{acknowledgments}

\bibliography{ref.bib}

\begin{thebibliography}{70}
\bibitem [{not()}]{noteLM}%
  \BibitemOpen
  \href@noop {} {}\bibinfo {note} {For $\alpha<1$, each site “sees” a
  mean-field $\propto L^{1-\alpha}$ which diverges with $L$, and $\sim\ln L$
  for $\alpha=1$.}\BibitemShut {Stop}%
\end{thebibliography}
\bibliographystyle{apsrev4-2}

\clearpage
\onecolumngrid

\begin{center}
	\textbf{\large Supplemental Material for \\"Unconventional Scalings of Quantum Entropies in Long-Range Heisenberg Chains"}
\end{center}
\setcounter{equation}{0}
\setcounter{figure}{0}
\setcounter{table}{0}
\setcounter{page}{1}
\setcounter{section}{0}

\makeatletter
\renewcommand{\theequation}{S\arabic{equation}}
\renewcommand{\thefigure}{S\arabic{figure}}
\setcounter{secnumdepth}{3}

\section{Extrapolation of $\ell^{P}_{2}(\alpha)$ for different $\alpha$}
\begin{figure}[htbp]
\centering
\includegraphics[width=0.84\columnwidth]{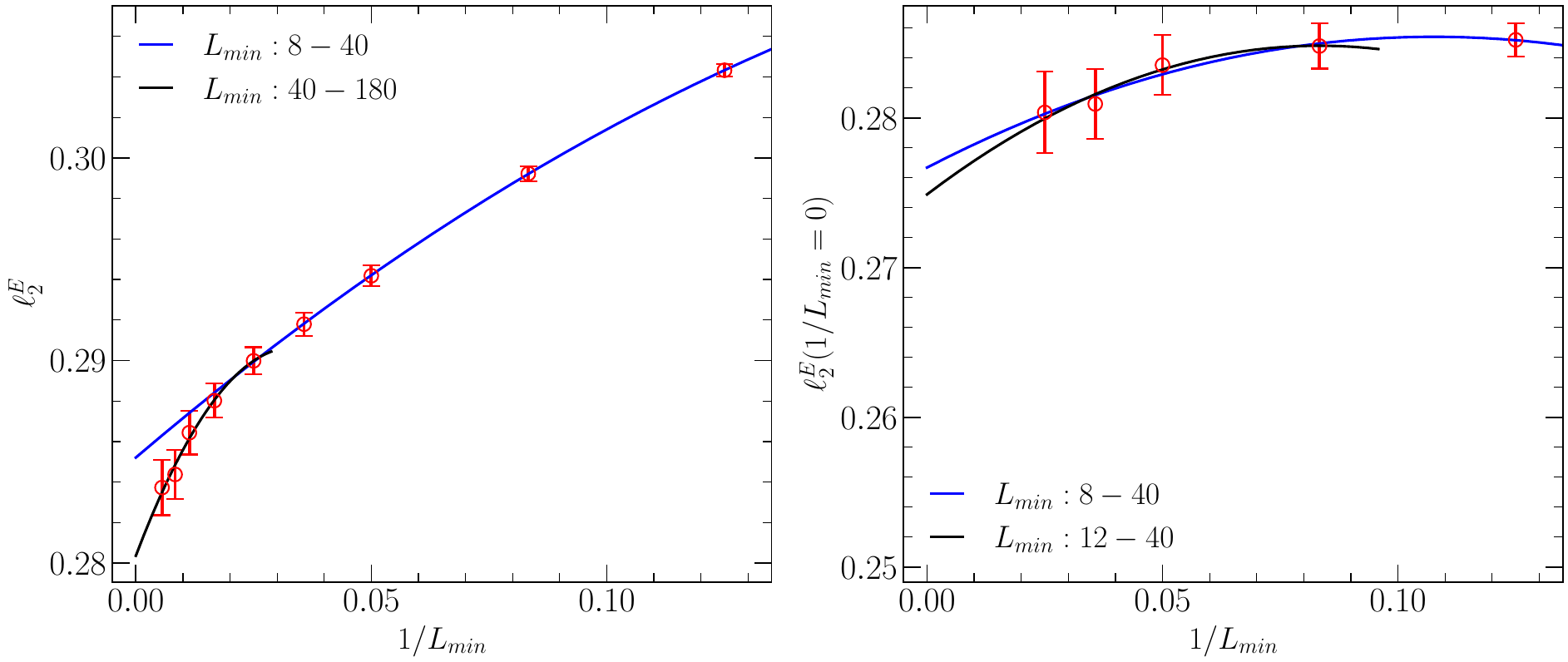}
\caption{\textbf{The finite size scaling of the $\ell_2^E(\alpha)$ at $\alpha_c=2.23$}. \textbf{Left panel:} The fitted $\ell_2^E$ versus $1/L_{min}$ for $\alpha=2.23$. Each data point $\ell_2^E(1/L_{min})$ is fitted from 6 consecutive finite-size EE data starting from $S^{E}_{2}(L_{\text{min}})$ with fitting function of $S^{E}_{2}=\ell_2^E \ln (L)+b$. The blue and black solid lines represent  extrapolating $\ell_2^E(1/L_{min}=0)$ with 5 consecutive finite-size data of $\ell_2^E(1/L_{min})$ with a quadratic function, starting from $L_{min}=8$(blue line) and $L_{min}=40$(black line) separately. \textbf{Right panel:} The fitted value $\ell_2^E(1/L_{min}=0)$, obtained in the left panel, versus $1/L_{min}$. The blue and black solid lines represent  extrapolating the converged  value of $\ell_2^E(1/L_{min}=0)$ with a quadratic function with all five(blue line) and the largest four(black line) data points included.} 
\label{fig:R1}
\end{figure}

\begin{figure}[htbp]
\centering
\includegraphics[width=0.84\columnwidth]{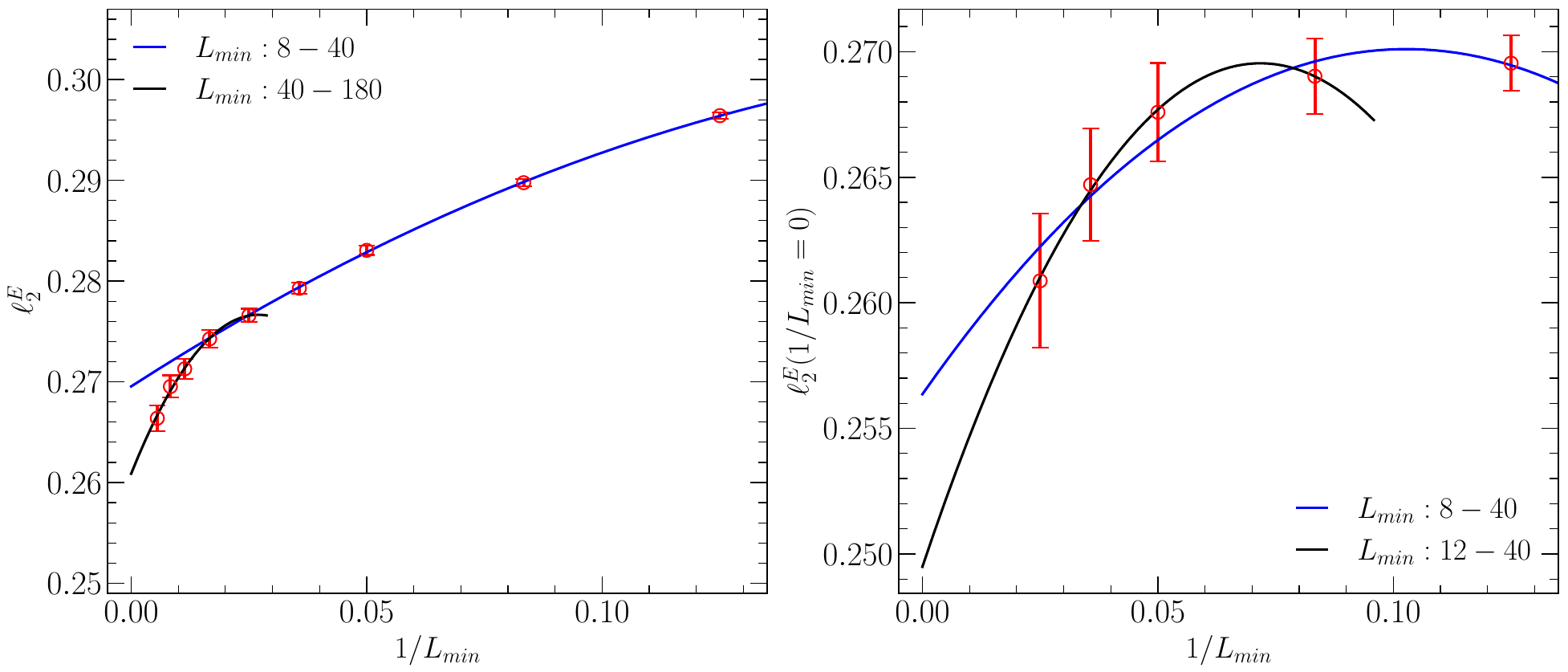}
\caption{\textbf{The finite size scaling of the $\ell_2^E(\alpha)$ at $\alpha=2.3$,  close to $\alpha_c$ while slightly lies in the short range regime}. \textbf{Left panel:} The fitted $\ell_2^E$ versus $1/L_{min}$ for $\alpha=2.3$. Each data point $\ell_2^E(1/L_{min})$ is fitted from 6 consecutive finite-size EE data starting from $S^{E}_{2}(L_{\text{min}})$ with fitting function of $S^{E}_{2}=\ell_2^E \ln (L)+b$. The blue and black solid lines represent  extrapolating  $\ell_2^E(1/L_{min}=0)$ with 5 consecutive finite-size data of $\ell_2^E(1/L_{min})$ with a quadratic function, starting from $L_{min}=8$(blue line) and $L_{min}=40$(black line) separately. \textbf{Right panel:} The fitted value $\ell_2^E(1/L_{min}=0)$, obtained in the left panel, versus $1/L_{min}$. The blue and black solid lines represent  extrapolating the converged  value of $\ell_2^E(1/L_{min}=0)$ with a quadratic function with all five(blue line) and the largest four(black line) data points included.}
\label{fig:R2}
\end{figure}

To explicitly understand the strong finite-size effect for entanglement entropy (EE) data, we use a more sophisticated method to  fit the thermodynamic value of $\ell_{2}^{E}$, with $\alpha=2.23$ and $\alpha=2.3$ as examples. The first step of this strategy is to gradually slide the fitting window, which contains 6(5) consecutive finite-size EE data for $\alpha>1$($\alpha=1$) starting from $S^{E}_{2}(L_{\text{min}})$, and apply the fitting function of $S^{E}_{2}=\ell_2^E \ln (L)+b$ to obtain the finite-size  $\ell_{2}^{E}(1/L_{min})$. The next step is to fitting the data points of $\ell_{2}^{E}(1/L_{min})$, as shown in the left panels of Figs.~\ref{fig:R1} and \ref{fig:R2}, to their thermodynamic limit $1/L_{min}=0$ to obtain the true value of $\ell_{2}^{E}$. However as shown in the the left panel of Figs.~\ref{fig:R1} and \ref{fig:R2}, when different sets of data points are used in the fitting process, the fitted value of $\ell_{2}^{E}(1/L_{min})$ is not always converged, which demonstrates strong finite-size effects.  We then further try to extrapolate the fitted $\ell_{2}^{E}(1/L_{min}=0)$, obtained from sliding the fitting window in the left panel, to study its convergence behavior. As shown in the right panels of Figs.~\ref{fig:R1} and \ref{fig:R2}, it is clear that for $\alpha=2.23$ the values of fitted $\ell_{2}^{E}(1/L_{min}=0)$ is well converged to a value around 0.28. While there is still finite size effect, the extrapolation with all five data points considered and the largest four data points considered separately give $\ell_{2}^{E}=0.277(4)$(blue line) and $\ell_{2}^{E}=0.275(9)$(black line), which are all significantly different with the short-range value of $\ell_{2}^{E}=0.25$. On the other hand, as shown in the right panel of Figs.~\ref{fig:R2}, for $\alpha=2.3$, the fitted $\ell_{2}^{E}$ exhibits a tendency to converge to the expected value of 0.25 for the short-ranged spin chain. The extrapolation with all five data points considered and the largest four data points considered separately give $\ell_{2}^{E}=0.256(4)$(blue line) and $\ell_{2}^{E}=0.249(9)$(black line). In practice, we choose $\ell_{2}^{E}=0.275(9)$ and $\ell_{2}^{E}=0.249(9)$ as our final inputs of the extrapolated thermodynamic results for $\alpha=2.23$ and $\alpha=2.3$ separately, because in the fittings of them the smallest system size is excluded so that they should have smaller finite-size effects.  

\section{Extrapolation of $\ell^{P}_{\infty}(\alpha)$ for different $\alpha$}
\begin{figure}[htbp]
\centering
\includegraphics[width=0.85\columnwidth]{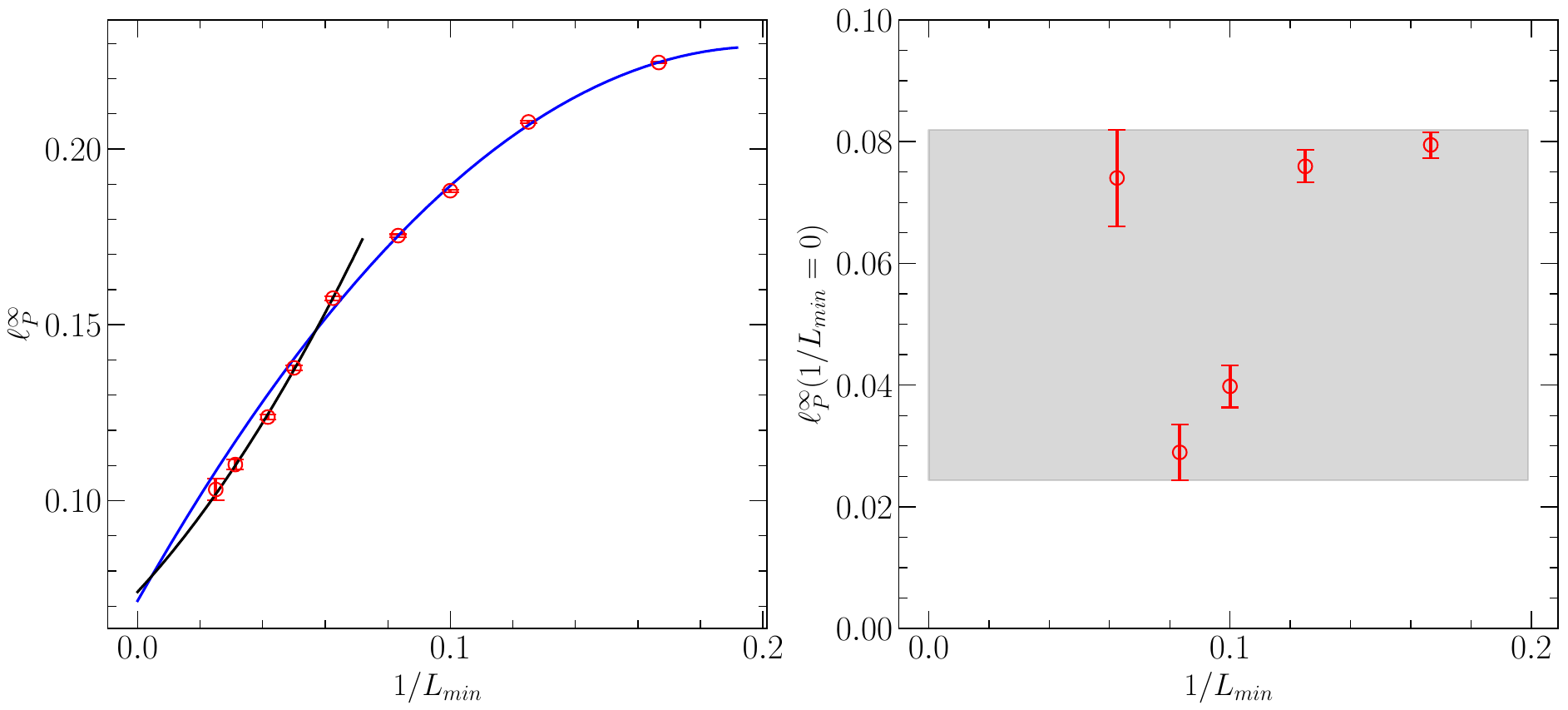}
\caption{\textbf{The finite size scaling of the $\ell_{P}^{\infty}$ at $\alpha_c=2.23$}. \textbf{Left panel:} The fitted $\ell_{P}^{\infty}$ versus $1/L_{min}$ for $\alpha=2.23$. Each data point $\ell^{\infty}_P(1/L_{min})$ is fitted from 6 consecutive finite-size PE data starting from $S^{P}_{\infty}(L_{\text{min}})$ with fitting function of $S^{P}_{\infty}=V_{P}^{\infty}L+\ell_{P}^{\infty} \ln (L)+b$. The blue and black solid lines represent  extrapolating $\ell_{P}^{\infty}(1/L_{min}=0)$ with 5 consecutive finite-size data of $\ell_{P}^{\infty}(1/L_{min})$ with a quadratic function, starting from $L_{min}=6$(blue line) and $L_{min}=16$(black line) separately. \textbf{Right panel:} The fitted value $\ell_{P}^{\infty}(1/L_{min}=0)$, obtained in the left panel, versus $1/L_{min}$. The shaded area refers to the fluctuation range of data points.} 
\label{fig:R3}
\end{figure}

\begin{figure}[htbp]
\centering
\includegraphics[width=0.85\columnwidth]{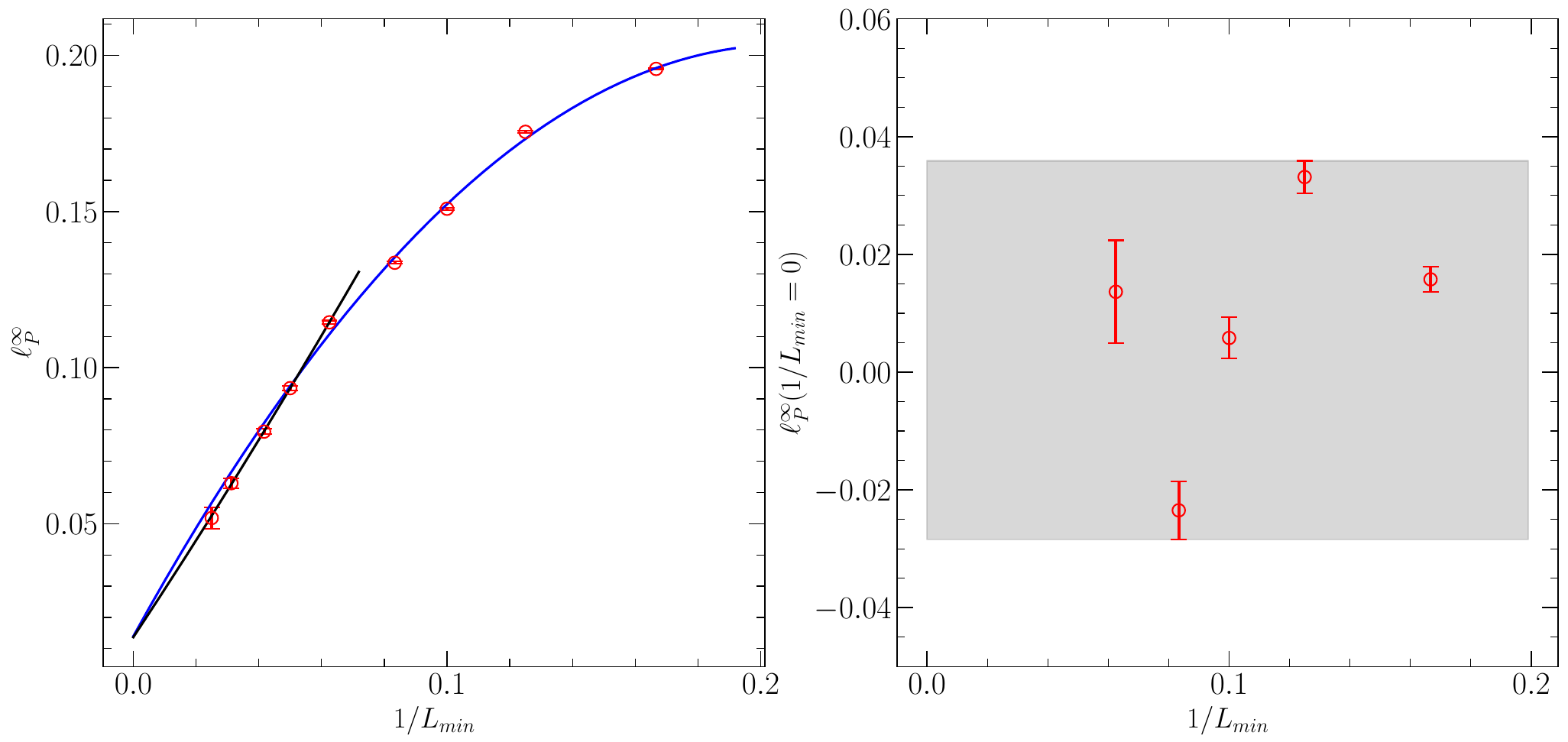}
\caption{\textbf{The finite size scaling of the $\ell_{P}^{\infty}$ at $\alpha=2.3$}. \textbf{Left panel:} The fitted $\ell_{P}^{\infty}$ versus $1/L_{min}$ for $\alpha=2.3$. Each data point $\ell^{\infty}_P(1/L_{min})$ is fitted from 6 consecutive finite-size PE data starting from $S^{P}_{\infty}(L_{\text{min}})$ with fitting function of $S^{P}_{\infty}=V_{P}^{\infty}L+\ell_{P}^{\infty} \ln (L)+b$. The blue and black solid lines represent  extrapolating $\ell_{P}^{\infty}(1/L_{min}=0)$ with 5 consecutive finite-size data of $\ell_{P}^{\infty}(1/L_{min})$ with a quadratic function, starting from $L_{min}=6$(blue line) and $L_{min}=16$(black line) separately. \textbf{Right panel:} The fitted value $\ell_{P}^{\infty}(1/L_{min}=0)$, obtained in the left panel, versus $1/L_{min}$. The shaded area refers to the fluctuation range of data points.} 
\label{fig:R4}
\end{figure}

In Fig.~\ref{fig:R3} and Fig.~\ref{fig:R4}, we apply similar data analysis procedure as we have done for EE data, while applying a different fitting function for finite-size data of participation entropy (PE) with $S^{P}_{\infty}=V_{P}^{\infty}L+\ell_{P}^{\infty} \ln (L)+b$. However, the difference from EE data is that, the finite-size dependence for PE data is not that strong as the EE data. In this case, as shown in the right panels of Fig.~\ref{fig:R3} and Fig.~\ref{fig:R4}, we observe fluctuations of fitted $\ell_{P}^{\infty}(1/L_{min}=0)$ around their mean values, rather than showing a clear dependence on $1/L_{min}$ as observed in EE data as shown in Fig.~\ref{fig:R1} and Fig.~\ref{fig:R2}. At $\alpha=2.23$, the fitted values have a mean value and uncertainty of $\ell_{P}^{\infty}=0.059\pm 0.021$, and At $\alpha=2.3$, the fitted values have a mean value and uncertainty of $\ell_{P}^{\infty}=0.009\pm 0.018$. 

\begin{figure}[htp!]
\includegraphics[width=\columnwidth]{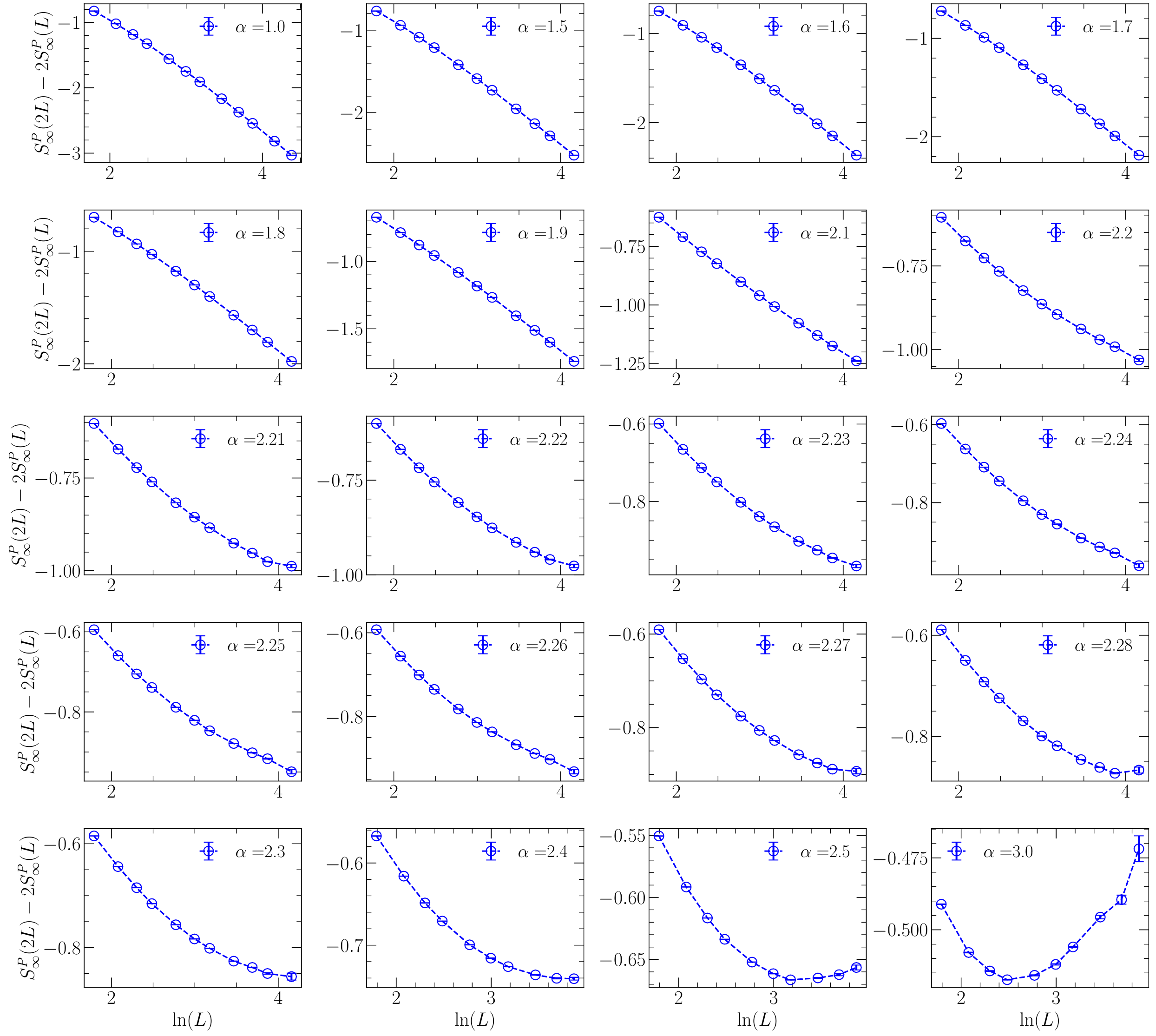}
\caption{$S^{{P}}_{\infty}(2L)-2S^{{P}}_{\infty}(L)$ versus $\ln (L)$ for different $\alpha$ which cancels out the leading area law and exposes the logarithmic corrections (if any) as leading term of the scaling form.} 
\label{fig:figs3}
\end{figure}

Additionally, as shown in Fig.~\ref{fig:figs3}, after eliminating the volume term in PE  by plotting $S^{{P}}_{\infty}(2L)-2S^{{P}}_{\infty}(L)$ against $\ln (L)$, at least when $\alpha\ge 2.27$, $S^{{P}}_{\infty}(2L)-2S^{{P}}_{\infty}(L)$ become flat and begins to grow as $L$ in increased. This shows that in the SR phase, $S^{{P}}_{\infty}(2L)-2S^{{P}}_{\infty}(L)$ is not linearly dependent on $\ln (L)$ anymore, but more likely to vary with $L$ due to finite size effects, consistent with our expectations. Furthermore, as in the short-range regime, $S_{\infty}^{P}=V_{\alpha}L-\ln\sqrt{K}+o(1)$ where $K$ is the LL parameters, we would expect as $L$ increases, $S^{{P}}_{\infty}(2L)-2S^{{P}}_{\infty}(L)$ will eventually converge to $\ln\sqrt{K}\sim -0.3466$.  Our data for $\alpha=3.0$ clearly shows this tendency, although is still not converged.
It is surely of great interest to push the system size to larger values to further examine this issue.

\end{document}